# IDENTIFYING ROTE LEARNING AND THE SUPPORTING EFFECTS OF HINTS IN DRILLS

G. Stefansson, A.H. Jonsdottir, T. Jonmundsson, G.S. Sigurdsson, I.L. Bergsdottir

*University of Iceland, Science Institute (ICELAND)*

## Abstract

Whenever students use any drilling system the question arises how much of their learning is meaningful learning, which emphasises understanding and the transferability of prior knowledge, and how much is memorisation through repetition or rote learning. Although both types of learning have their place in an educational system it is important to be able to distinguish between these two approaches to learning and identify options which can dislodge students from rote learning and motivate them towards meaningful learning.

The tutor-web is an online drilling system, which has been used by thousands of students from Iceland to Kenya. The design aim of the system is to promote meaningful learning rather than evaluation. This is done by presenting students with multiple-choice questions which are selected randomly but nevertheless linked to the students' performance to ensure that students are appropriately challenged. The questions themselves can be generated for a specific topic by drawing correct and incorrect answers from a collection associated with a general problem statement or heading. With this generating process students may see the same question heading twice but be presented with all new answer options or a mixture of new and old answer options.

Data from an introductory university course on probability theory and statistics, taught using the tutor-web during COVID-19, are analysed to separate rote learning from meaningful learning. The analyses show that considerable non-rote learning takes place, but even with fairly large question databases, students' performance is considerably better when they are presented with an answer option they have seen before. An element of rote learning is thus clearly exhibited but a deeper learning is also demonstrated.

The item database has been seeded with occasional hints such that some questions contain fairly detailed clues, which should cue the students towards the correct answer. This ties in with the issue of meaningful learning versus rote learning since the hope is that a new hint will work as a cue to coax the student to think harder about the question rather than continue to employ rote learning. The existence of occasional hints allows several comparisons. The simplest analysis is on whether the overall grade on cue questions is higher than on the non-cue questions. A more important issue is whether more learning has occurred and methods are developed to estimate the change rather than status. Preliminary results indicate that hints are particularly useful for students with poor performance metrics, and a power analysis demonstrates the sample sizes needed in future studies to better quantify these effects.

## 1 INTRODUCTION

The tutor-web [1] system is designed for research [2] and learning [3]. Its drills are primarily used for learning so there are typically no limits on the number of attempts at improving performance. An important feature is that for most drills the student is shown a detailed explanation of the solution immediately after choosing an answer option. This system is used at multiple schools and universities in Iceland and Kenya, mostly for mathematics and statistics. Students earn SmileyCoin, a cryptocurrency, while studying [4].

The design aim of the system is to promote meaningful learning rather than evaluation. This is implemented by presenting students with multiple-choice questions, which are selected randomly, but linked to the students' performance to ensure that students are appropriately challenged. Other project types have also been used in the tutor-web but are not the topic of the present paper.

In the experience of the authors many students do not purchase textbooks and many university students do not appear to have learnt in secondary school the art of continually following the thread of communication from the instructor. It appears that a fairly large proportion of students only put hours into studying when they have an assignment due. Assuming that there is no easy way to convince such students to read the assigned material, one is led to ask whether the reading or knowledge assimilation function can be included as a natural part of assignments, even for automated on-line drills.

For most items in the tutor-web drill database, a solution description has been set up and this is shown to the student after they have submitted an answer to the drill question. This was initially done in response to student requests for improvement but it is certainly found useful by enthusiastic and diligent students. However, it is also seen that many students do **not** read these after-the-fact explanations but merely continue to request and answer new items as if on autopilot. A different mechanism is therefore needed to increase learning, since it seems the learning needs to occur during the act of solving the problem.

This issue has been studied in different context by several authors, for example in physics education in the form of **hints** in a student-controlled problem-solving program [5], in English education with contextual **clues**, which are put in texts to support student's understanding [6] and in teaching biology with **cues** provided to students before assigning or discussing questions [7]. The differences between hints, clues and cues are subtle but each may be used to increase understanding and problem-solving capability as a part of an assignment: A clue may be a hidden part of a question which the student needs to read carefully to find it, whereas a hint would normally be a more explicit expression of a direction. In either case the intent is for the support to work as a cue so that the student directs attention towards working out the problem.

## 2   METHODOLOGY

### 2.1.1   Designing drill sets

The data analysed to distinguish meaningful learning consist of student responses to drills in an introductory university course on probability theory and statistics, taught using the tutor-web during COVID-19. Student evaluation in the course consisted almost exclusively of these drills as no in-house finals or mid-terms were allowed once classrooms were closed due to the pandemic. The students initially were given a handful of drills as homework, but subsequently the drill sets were expanded and new drills were generated for use in a mid-term and as a final exam. Several components of the drill sets were re-used in subsequent exams.

The most common method to design tutor-web drills is to set up collections of multiple-choice items in the form of a drill set outside the system and subsequently upload the entire drill set. A drill set is commonly a collection of drills on a fairly narrow topic, but can in principle also be an overview collection on several topics.

The three main approaches to designing drill sets for the tutor-web are (a) handcrafting individual items (b) using random numbers to generate an entire drill set based on a single item and (c) use a generic "check the appropriate answer" header with a choice of a correct option and several distractors, where both the correct answer and distractors are chosen randomly from a reasonably large collection of possible options.

The mid-term had four drill sets. Each was composed of 300 drill items based on a random selection of correct options and distractors with occasional "None of the Above" or "All of the Above", which could be either correct or incorrect (NOTA+/NOTA-/AOTA+/AOTA-). The underlying number of correct choices and distractors in each drill set were (20, 48), (13, 13), (41, 58), (40, 52).

For the final exam, eight drill sets with a total of 2380 drill items were constructed. Some of these sets were based on earlier homework or the mid-terms, but some were completely new. Each drill set had a single header giving an introduction to the problem, followed by "check the most appropriate box". All the items were generated using approach (c) above, by choosing a correct option and distractors from a set, possibly appending a NOTA or AOTA, each of which could be correct or incorrect (with appropriate probability). When thousands of items are needed as in the introductory stats course, the first approach was not feasible and it was found that, given the time available, approach (c) was the most feasible option. The total number of distractors was chosen randomly using a truncated Poisson distribution, except in the NOTA/AOTA cases, where the NOTA/AOTA option was always the fourth and last option. The composition of the drill sets used for the final exam is given in Table 1.

*Table 1. Final exam: Number of underlying options and number of generated items.*

| Drillset | Correct Set | Distractor Set | Generated items |
| --- | --- | --- | --- |
| 1 | 43 | 60 | 280 |
| 2 | 41 | 53 | 300 |
| 3 | 15 | 24 | 300 |
| 4 | 13 | 23 | 300 |
| 5 | 45 | 62 | 300 |
| 6 | 16 | 38 | 300 |
| 7 | 26 | 35 | 300 |
| 8 | 24 | 38 | 300 |
| Total | 223 | 333 | 2380 |

Although the tutor-web system does try to increase the difficulty of the questions as the grade increases, this effect should be minor when the topics are as narrow as they are here.

Aside from this particular statistics course, the full item database of tutor-web has been seeded with occasional hints so that some questions contain fairly detailed clues, which should cue the students towards the correct answer. The level of detail varies from subtle hints to an extensive description of how to solve the problem. This ties in with the issue of meaningful learning versus rote learning since the hope is that a new hint will work as a cue to coax the student to think harder about the question rather than continue to employ rote learning. The existence of occasional hints allows several comparisons. The simplest analysis is on whether the overall grade on cue questions is higher than on the non-cue questions. A more important issue is whether more learning has taken place and methods are developed in this paper to estimate the change rather than status.

*2.1.2 Data preparation and analysis of responses to identify rote learning*

As the student progresses through a drill set, they obtain a 0/1 grade ($g_t$) at each sequential step (t), where the coding is 1=correct, 0=incorrect. The fact that these grades tend to be increasing is well-known [1] and this does reflect some sort of learning, but the task here is to distinguish between rote learning and a deeper understanding.

Since the tutor-web assigns drills randomly *with replacement* from a pool, students will occasionally see the exact same item again, but in the intro stats course, many items had been generated using method (c) above. Each time a student sees an item of this type, the student may in principle have seen the exact same item before, but to investigate rote learning it is of greater importance to evaluate whether the specific correct answer option had been seen before. To separate rote learning from meaningful learning, student responses to each item were therefore classified according to whether the correct answer option had been seen before or not.

*2.1.3 Analysis of responses to identify the role of hints*

The data analysed to identify the effects of including cues or hints are all available responses in the tutor-web corresponding to pairs of items, where one has a hint and the other does not. There are 231 such item pairs in the data base and the students have seen and answered such items 30,352 times.

Basically, if $g_t$ is the student's 0/1-performance grade for the t'th item in a drill sequence, the simplest analysis is checking whether $g_t$ is on average greater for the cue questions but a more advanced analysis investigates **the learning due to seeing an item,** which is the change in grade across the t'th item: $g_{t+1}-g_{t-1}$ and evaluate whether these differences are on average greater if the t'th item has some type of clue. It is also important to verify whether the clue has a differential effect depending on the students existing knowledge and whether the effects are different depending on the specific drill.

For this analysis to be meaningful, the paired items need to be similar in the sense that they are all from a collection of almost identical items except for the occasional clue.

## 3 RESULTS

### 3.1 Rote learning or not

#### 3.1.1 Basic results

As mentioned above, each single drill set consists of questions on a fairly narrow topic, generated from a total of 16-45 truly different correct answers with added distractors. Earlier results [1,3] have used the collection of answers to demonstrate how students' grades tend to increase with the number of attempts in a drill set. One problem with using this approach on tutor-web data is that the students see some of the correct choices repeatedly or even the exact same drill several times.

A better decomposition of learning is seen in Fig. 1. In this figure, the x-axis denotes the sequential attempts (t) of a student trying their way through a drill set and the y-axis denotes the average proportion of correct answers. The red dots in the picture show the average grade obtained by a student as they are given previously unseen correct answers.

The red curve is a simple model fit to the red points. The increase in the fitted curve clearly demonstrates an increase in knowledge on this narrow topic: The average grade **on previously unseen correct answers** increases as the students continue to work on the problems.

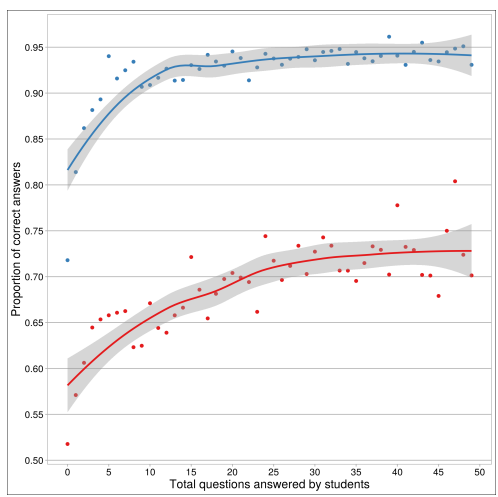

*Figure 1. Performance increase as a function of the number of attempts. The number of attempts is within a given topic (drill set) and truncated to 50 trials. The performance (y-axis) is measured as the proportion of correct answers within the drill set. The data are classified according to whether the student has seen the correct response before (blue dots) or not (red dots). Also shown are smoothed curves with shaded error regions.*

Conversely, the blue dots show the average grade as a function of the number of attempts, but only on questions that the student has seen before. The difference between the two curves is a clear indication that there is considerable rote learning, **which does not map into increased understanding or transferability new and previously unseen problems**. The uptake of deeper understanding and transferability (red curve) is much slower than the rote learning but nonetheless very clear.

The way they stand, these results are of course merely qualitative, but formal models of these data can be used to demonstrate that the effects are real and highly significant. A binomial/logistic response model was set up for this purpose. The model included an indicator variable for whether the response had been seen before as well as the number of attempts, to give the output needed for Fig. 1. In addition, other variables were inserted to correct for differences among students (random effect), topic, as well as measures of increasing difficulty of questions and the number of distractors. All of the fixed effects in the model were highly significant, justifying the above interpretations of Fig. 1.

### 3.1.2 The effect of seeing repeated distractors

In addition to having seen the correct answer, it should be noted that when using the above method of generating drill items, it may also make a difference whether the students have seen one or most of the distractors before. This is particularly the case when the distractor set is small. This is not really the case in the present data set, but distractors can be difficult to generate so this is potentially an important issue.

The above logistic model was augmented using the proportion of distractors seen before when a student was given a new drill item. It turns out that having seen the distractors before is highly significant and the earlier parameters remain statistically significant.

## 3.2 The effects of clues

### 3.2.1 Basic results for struggling students

At first sight, the overall direct effect of including a hint/clue or not appears to be quite small, since the average grade on a regular question is 72.7% but 77.4% on a cue-question.

However, as is so often the case, the devil is in the detail. It is particularly important to see whether a hint helps the struggling student. By splitting the student body into two groups, according to whether the existing grade was over or under 50%, one finds that the struggling group receives an average grade on 53.1% (n=10,809) on non-cue questions but 60.4% (n=2,797) if the question has a cue, i.e. an increase of 6.3%. The difference is much smaller for the groups with more than 50% initially, where the grade is 89.9% (n=12,341) on the non-cue questions but 91.3% (n=3,406) on the cue questions, so there is still an increase but only of 1.4%.

These preliminary results indicate that hints are particularly useful for students with poor performance metrics.

### 3.2.2 Variations among drills

Not surprisingly, the effect of giving a hint is also quite variable depending on the particular question: This is seen by fitting a binomial response model to the data, including a two-way analysis of variance with interaction, with one factor for the question and the other describing if it had a hint or not. The students were taken into account as a random effect. The net result is that the interaction term is highly significant and also explains a nontrivial portion of the deviance.

### 3.2.3 Measures of actual learning from receiving a cue

Finally, consider a simple measure of actual learning, in the form of a change in item grade, $g_{t+1}-g_{t-1}$, across having obtained a cue at time t but not at time t-1 or t+1.

As the student works through a drill set a sequence of 0/1 data is obtained. For each such grade measurement, $g_t$, a whole suite of other information is also recorded. In particular one can add information on whether the drill had some form of hint and use this as an indicator variable which we will call *cue* or $c_t$. The student might in principle see many questions with cues throughout the drill set, and the first such might appear at any time. Thus, there are several options on how to consider the relationship between $g_{t+1}-g_{t-1}$ and the indicator $c_t$ for a hint.

To reduce confounding only the student's first triplet of answers from a drill set was used and those triplets selected where the cue was only on the second question. Thus, the data consist of those numbers, $g_3-g_1$, where $c_1=0$, $c_2=1$, $c_3=0$. A summary of these data is given in Table 2.

*Table 2. Frequencies of changes in grade (learning, $g_{t+1}-g_{t-1}$), depending on whether a cue was given in between or not, based on all available starting triplets of answers from all drill sets in the tutor-web.*

| Learning (grade change) | No cue | Cue | Total |
|---|---|---|---|
| -1 | 93 | 8 | 101 |
| 0 | 1014 | 53 | 1067 |
| 1 | 601 | 41 | 641 |
| Total | 1708 | 102 | 1809 |

The simplest summary of the table is to investigate the proportion of learning ($g_{t+1}-g_{t-1}$=1), which is 35.1% for the non-cue column vs 40.2% for the cue column. The significance of this difference is somewhat difficult to evaluate given the correlated data, but it is not significant under the assumption of independent measurements, so more complex models are unlikely to give statistical significance. A particular difficulty with these data is that many of the learners will always obtain a perfect grade so their change will always be zero.

Next, consider only those students in the above table who answered the first item incorrectly. Some of them next got a hint but others did not.

This time there are only 805 responses and only 47 received a hint as a part of the second item. Without a hint, 79% answered the third item correctly but 87.2% answered correctly after a hint on the second item. As before these numbers are not statistically significant (p=0.26 assuming independent counts).

It follows that although these initial results are extremely promising, more data needs to be gathered to obtain a better estimate of the effect of including hints in occasional drill items. It would also be useful to see the different effect of hints on learning depending on whether the questions were in a very tight group with a single question header (as in the stats course) or a more general group such as questions on a typical lecture or lecture collection (as was the case in this analysis).

### 3.2.4 Data requirements for further research

Although more extensive data are needed to draw firm conclusions from this kind of analysis, it does seem that the effects of hints in questions can be quite important, particularly for the struggling student, yet highly dependent on the question chosen. These results from elementary data analysis hold up through simple models and give reason enough to consider extensive use of several types of cues in drills to see how and when they can be used to assist struggling students.

A power analysis demonstrates the sample sizes needed in future studies to better quantify these effects. Addressing the above frequency-table analysis directly as a power estimation exercise is easily done using a simple Monte Carlo sampling of the lines in the data matrix and evaluating the probability of rejection by the chi-square test. By varying the number of sampled data it is found that if the 805 responses could be increased to 3450 then there is an 80% chance of obtaining a significant result if the same structure holds in the data.

## 4 CONCLUSIONS

### 4.1.1 Measuring understanding above rote learning

The above analyses show that considerable non-rote learning takes place, but even with these fairly large question databases, students' performance is considerably better when they are presented with an answer option they have seen before.

An element of rote learning is therefore clearly exhibited but a deeper learning is also demonstrated.

These results affect choices of item generation in drilling systems. As mentioned above, there are three common methods to generate items for the tutor-web: (a) hand-crafting, (b) using random numbers in a single format and (c) randomly sampling a correct item with several distractors. A fourth approach can be implemented as a combination of (b) and (c), given enough resources. For example, to teach students how to interpret regression output, option (c) would present static regression output and ask dozens of questions such as "is the estimated slope significantly different from zero" or "what percentage of the variation in the data is explained by the model". The combined approach would be to randomly generate data, run the regression and choose one of the random items, but based on the generated data. It would be expected that the combined approach should be better than (c) alone, but requires considerably greater time (and dedicated programming) to set up. This type of approach has been used in [8] where real data sets are mined to extract random subsets to be used in random questions.

*4.1.2   The importance of hints in drills*

Preliminary results indicate that hints are particularly useful for students with poor performance metrics and should be routinely included, not only to enhance the performance of these students but also to better elucidate where these interventions are most useful.

A power analysis demonstrates the sample sizes needed in future studies to better quantify these effects. These sample sizes could be obtained by including cues in five times as many samples as have been obtained to date, which is quite feasible.

## ACKNOWLEDGEMENTS

A large number of individuals and institutions have made this work possible. Through the years, the projects have received funding from The Icelandic Centre for Research and from several EU H2020 grants, most recently FarFish (Horizon 2020 Framework Programme Project: 727891 — FarFish).

Continuous support has been provided by the University of Iceland where the course material has been developed, and by the University of Iceland Science Institute where most of the research and development has been conducted. The current version of the tutor-web was developed by Jamie Lentin at Shuttle Thread Ltd and Jamie has also looked participated in the development of the SmileyCoin wallets.

Countless students have contributed to the tutor-web and the SmileyCoin wallet, most recently Jóhann Haraldsson who led the TA group for the intro stats course in the middle of COVID-19.